\documentclass[usenatbib]{mnras}

\usepackage{multicol}
\usepackage{aas_macros,times,multirow,amsmath,amssymb,longtable,breqn}

\usepackage[T1]{fontenc}
\usepackage[varg]{txfonts}

\usepackage{grffile} 
\usepackage{graphicx}

\usepackage{color}
\usepackage{xcolor}
\usepackage{ulem}

\newcommand\U[1]{{\,\rm #1}}

\newcommand\ltsima{$\; \buildrel < \over \sim \;$}
\newcommand\lsim{\lower.5ex\hbox{\ltsima}}
\newcommand\gtsima{$\; \buildrel > \over \sim \;$}
\newcommand\gsim{\lower.5ex\hbox{\gtsima}}

\newcommand{\be}{\begin{equation}}
\newcommand{\en}{\end{equation}}

\newcommand\rs[1]{_\mathrm{#1}}
\newcommand\Esn{E\rs{sn}}

\newcommand\rhoej{\rho\rs{ej}}
\newcommand\vej{v\rs{ej}}
\newcommand\Pej{P\rs{ej}}
\newcommand\Mej{M\rs{ej}}
\newcommand\Rforw{R\rs{fs}}
\newcommand\dotRforw{\dot{R}\rs{fs}}
\newcommand\Rrev{R\rs{rs}}
\newcommand\Rpwn{R\rs{pwn}}
\newcommand\vpwn{v\rs{pwn}}
\newcommand\Ppwn{P\rs{pwn}}
\newcommand\vt{v\rs{t}}
\newcommand\rhoism{\rho\rs{ism}}
\newcommand\nism{n\rs{ism}}

\newcommand\Rmax{R\rs{max}}
\newcommand\Rmin{R\rs{min}}
\newcommand\Rch{R\rs{ch}}
\newcommand\tch{t\rs{ch}}

\begin{document}
\label{firstpage}
\pagerange{\pageref{firstpage}--\pageref{lastpage}}

\title[Reverberation of PWNe (I)]{Reverberation of pulsar wind nebulae (I): \\
Impact of the medium properties and other parameters upon the extent of the compression}
\author[Bandiera et al.]
{R. Bandiera$^{1}$, N. Bucciantini$^{1,2,3}$, J. Mart\'in$^{4,5}$, 
B. Olmi$^{1,4}$, D. F. Torres$^{4,5,6}$ \thanks{All authors have contributed equally to this work.}  \\
$^{1}$ INAF - Osservatorio Astrofisico di Arcetri, Largo E. Fermi 5, I-50125 Firenze, Italy \\
$^{2}$ Dipartamento de Fisica e Astronomia, Universit\`a degli Studi di Firenze, Via G. Sansone 1, I-50019 Sesto F. no (Firenze), Italy \\
$^{3}$ INFN - Sezione di Firenze, Via G. Sansone 1, I-50019 Sesto F. no (Firenze), Italy \\
$^{4}$ Institute of Space Sciences (ICE, CSIC), Campus UAB, Carrer de Can Magrans s/n, 08193 Barcelona, Spain \\
$^{5}$ Institut d'Estudis Espacials de Catalunya (IEEC), Gran Capit\`a 2-4, 08034 Barcelona, Spain \\
$^{6}$ Instituci\'o Catalana de Recerca i Estudis Avan\c cats (ICREA), 08010 Barcelona, Spain 
}

\date{}
\maketitle

\pubyear{2019}

\begin{abstract}

The standard approach to the long term evolution of pulsar wind nebulae (PWNe) is based on one-zone models treating the nebula as a uniform system.
In particular for the late phase of evolved systems, many of the generally used prescriptions are based on educated guesses for which a proper assessment lacks. 
Using an advanced radiative code we evaluate the systematic impact of various parameters, like the properties of the supernova ejecta, of the inner pulsar, as well of the ambient medium, upon the extent of the reverberation phase of PWNe.
We investigate how different prescriptions shift the starting time of the reverberation phase, how this affects the amount of the compression, and how much of this can be ascribable to the radiation processes.
Some critical aspects are the description of the reverse shock evolution, the efficiency by which at later times material from the ejecta accretes onto the swept-up shell around the PWN, and finally the density, velocity and pressure profiles in the surrounding supernova remnant.
We have explicitly treated the cases of the Crab Nebula, and of J1834.9--0846, taken to be representatives of the more and the less energetic pulsars, respectively.
Especially for the latter object the prediction of large compression factors is confirmed, even larger in the presence of radiative losses, also confirming our former prediction of periods of super-efficiency during the reverberation phase of some PWNe.

\end{abstract}

\begin{keywords}
radiation mechanisms: non-thermal -- pulsar: general -- method: numerical -- ISM: supernova remnants  
\end{keywords}

\section{Introduction}
\label{sec:intro}

Pulsar Wind Nebulae (PWNe), or plerions, are one of the most important particle accelerators in the Universe and the largest class of identified, Galactic very-high-energy gamma-ray sources \cite{HESS-GPS-2018}.
In PWNe, the wind of a young and active pulsar feeds a relativistic bubble, composed by magnetic fields and relativistic particles (believed to be mostly electron-positron pairs), and confined by the hosting Supernova Remnant (SNR).
So far those systems have been described with one-zone models, or equivalently 0+1 codes \citep{Bucciantini:2013a}, where the PWN is represented as a uniform system interacting with the SNR, subjected to adiabatic and radiative losses and possibly escape of particles.
One-zone models have been developed on top of the original work by \citet{Pacini:1973}, then extended with the description of the interaction with the SNR presented by \citet{Reynolds_Chevalier84a}.

From observations and simulations, we know that a young PWN expands, by slightly accelerating, inside the freely expanding SN ejecta, and in this way sweeping it
up to form a shell of material. 
This phase was largely simulated with hydrodynamic (HD) and magneto-hydrodynamic (MHD) codes, both in the classic and relativistic regimes, from 1D to 3D \citep{van-der-Swaluw:2001, Bucciantini:2003, Komissarov:2004, Del-Zanna:2006, Porth:2014, Olmi:2016}.
These series of works shows that one-zone models appear to be robust to describe that phase, with their assumption being valid.
This first phase proceeds until it reaches the reverse shock (RS, e.g. \citealt{Gaensler2006,Torres2017b}).
After that time, due to the mass accretion as well as to the thermal pressure of the shocked SNR medium, the shell experiences a strong deceleration, which in most cases leads to a compression of the PWN.
Afterwards, when due to compression the PWN internal pressure becomes high enough, the PWN bounces and re-expands again. 
This phase of contraction and re-expansion of the nebula bubble is called reverberation.
Only a few dynamical models had been extended to this late phase, and were mostly targeted to single objects do to their complexity, and they did not account for the study of a large set of physical parameters \citep{van-der-Swaluw:2001, Blondin:2001,Bucciantini:2003, Kolb:2017}.

The compression during the reverberation phase heats the particles and may cause a strong enhancement of the PWN magnetic field, in which case it would burn most of the high-energy electron-positron pairs that produce the characteristic multifrequency spectrum, possibly modifying the spectrum significantly.
HD and MHD models cannot account for the spectral evolution, for which only 0+1 codes can be use \citep{Bednarek:2003,Tanaka2010,Gelfand2009,Bucciantini2011,Vorster2013,Martin2016}.
For instance, the synchrotron burning could be so high, that periods in which the luminosity in X-rays or other frequencies exceeds the spin-down power can be found (see \citealt{Torres2018,Torres2019}).

However, despite reverberation is a critical phase in the evolution of PWNe, due to the many complexities it entails it is usually simplified (when not plainly neglected) in dynamical/radiative models. 
Aspects of the interaction between the RS and PWN shell, the mass loading at the shell, asymmetries in the collision between the aforementioned shells caused by irregularities in the ISM or in the SNR expansion, or instabilities after the PWN compression due to differences in the PWN/SNR ejecta densities (Rayleigh-Taylor like instabilities \citealt{Blondin:2001, Bucciantini:2004}), are very sensitive to some of the environmental parameters (as we discuss below) and are hardly well-described in full by a set of simplifying assumptions. 
Pulsars are also characterized by high-kick velocities at birth,  with average value of $\sim 350$ km/s \citep{Faucher2006}. Then the PSR may be displaced from the SNR center, making the interaction with the RS highly asymmetric also for geometrical reasons.
Ultimately, only detailed multi-dimensional HD models can account for these issues \citep{van-der-Swaluw:2003, Temim2015, Kolb:2017,Barkov2018, Olmi:2019, Olmi:2019a}.

Thus, a deeper understanding and modelling of the reverberation phase is a key issue in order to characterize the PWN population and  correctly determine parameters such as the age, the pair distribution spectrum, and the photon spectrum, at all subsequent ages beyond.
This would be particularly relevant in connection with future gamma-ray instruments, such as the Cherenkov Telescope Array (CTA), being PWNe the expected  dominant sources at gamma-rays \citep{de-Ona-Wilhelmi:2013,Klepser:2013,H.E.S.S.-Collaboration:2018a}.

This work is the first in a series of papers reporting the results of a research program that aims at doing exactly that: study reverberation in depth, from all angles, using different simulation tools.
Our aim here is to observe and quantify the impact of the medium properties and other parameters and assumptions upon the extent of the compression of the reverberation phase.
A common assumption for the reverberation phase is to take the bounding SNR in a fully relaxed Sedov state, following the Sedov solutions \citep{Sedov1959}.
Although, as we will explain below, this is not the case at the typical times of the reverberation because Sedov solutions are an accurate description only at later times, we will use them for our investigation of the parameters space.
In this paper we shall investigate, for instance, how deviations from the Sedov solution in terms of pressure and density of the SNR affect the results for the PWN-SNR interaction, and to estimate how robust these results are.
In forthcoming papers in the series we shall relate these radiative models results with hydrodynamical simulations, ultimately providing a prescription to go through this critical period of evolution.

\section{Dynamic equations in the {\sc TIDE} code} \label{sec:dyneq}

{\sc TIDE} \citep{Martin2012,Torres2014,Martin2016} is a time-dependent radiative 0+1 code that evolves the electron-positron population inside a PWN.
It does it by solving the diffusion-loss equation taking into account adiabatic losses, and synchrotron, inverse Compton (including self-Compton), and Bremsstrahlung radiative losses.

In this 0+1 code the hydrodynamic part is sketched in the following way: the PWN is a homogeneous bubble that, in its expansion, sweeps out ejecta material to form a thin, massive shell at its boundary.
As for the dynamics of this shell, our code follows similar treatments present in the literature (i.e., \citealt{Reynolds1984,Gelfand2009}).
The evolution of this massive shell, whose size is equal to the PWN radius $\Rpwn$, which in the following we shall simply call $R(t)$, is derived by solving the following set of equations
\begin{eqnarray}
\frac{dM(t)}{dt}\qquad & = &4 \upi R^2(t) \rhoej(R,t)  [v(t)-\vej(R,t)]\,;
\label{eqmass} \\
\frac{d}{dt}[M(t)v(t)] & = & F(t),
\label{eqmomentum}
\end{eqnarray}
where the first equation describes the mass conservation, and the second one does it with the momentum conservation of the shell. 
$v(t)=\dot R(t)$ is the shell radial velocity (also referred to as the PWN velocity, $\vpwn$), while $M(t)$ is the shell mass.
The force $F(t)$ is given by two components: the difference between the PWN and the ejecta pressure, and the contribution to the change of momentum due to the addition of SNR ejecta material. This yields
\begin{equation}
\label{force}
F(t)=4 \upi R^2(t)[\Ppwn(t)-\Pej(R,t)]+\frac{dM(t)}{dt}\vej(R,t)\,.
\end{equation}
The quantities $\rhoej(R,t)$, $\vej(R,t)$ and $\Pej(R,t)$, introduced above, are respectively the density, radial velocity and thermal pressure in the thermal SNR ejecta.
The quantity $\Ppwn(t)$, instead, is the pressure at the boundary of the PWN and is the sum of the pressure of the magnetic field and the one produced by the relativistic particles.
Modelling the PWN as a homogeneous relativistic bubble, the magnetic pressure ($P_B$) evolves according to the following equation.
\begin{equation}
\frac{d}{dt}(4 \upi R^3(t)P_B(t)) = \eta L(t)-4 \upi R^2(t)P_B(t)\frac{dR(t)}{dt}\,,
\label{eqPBevol}
\end{equation}
where the first term on the right-hand side gives the energy input (with an efficiency $\eta$) from the pulsar spin-down power
\begin{equation}
L(t) = L_0\left(1+\frac{t}{\tau_0}\right)^{-\frac{n+1}{n-1}}\,,
\end{equation}
being $L_0$ and $\tau_0$ are the initial power and spin-down time, while $n$ is the pulsar braking index.
The latter term on the right-hand side of Eq.~\ref{eqPBevol} accounts instead for the adiabatic losses.
As for the pressure of the particles component, in TIDE it comes out from the detailed calculation of the evolution of the particles energy distribution.
When radiative losses are negligible, this pressure evolution could be also described by an equation similar to Eq.~\ref{eqPBevol}, where now the efficiency for the pulsar input is $1-\eta$; but when radiative losses are energetically important a detailed calculation of the particles energy distribution becomes necessary, what the {\sc TIDE}
code does.

By substituting the mass derivative with the use of Eq.~\ref{eqmass}, one finally gets
\begin{eqnarray}
F(t)&=&4 \upi R^2(t)\left[\Ppwn(t)-\Pej(R,t)\right.\nonumber\\
&&\qquad\left.+\rhoej(R,t)\vej(R,t)\left(v(t)-\vej(R,t)\right)\right].\qquad
\label{eqFdef}
\end{eqnarray}
Expanding the derivative of the left term of Eq.~\ref{eqmomentum}, and using for the force its explicit expression (Eq.~\ref{eqFdef}), one then obtains
\begin{eqnarray}
M(t)\frac{dv(t)}{dt}&=&4 \upi R^2(t)\left[\Ppwn(t)-\Pej(R,t)\right.	\nonumber\\
&&\qquad\left.-\rhoej(R,t)\left(v(t)-\vej(R,t)\right)^2\right].\qquad
\label{eqvel}
\end{eqnarray}
The last term in the brackets takes the form of a ram pressure; however, it must be clear that it is present only by virtue of the fact that new material is accreted into the shell; while in the absence of such accretion Eqs.~\ref{eqmass} and \ref{eqvel} simplify to
\begin{eqnarray}
\frac{dM(t)}{dt} & = & 0;\\
\label{eqvel22}\
M(t)\frac{dv(t)}{dt} & =& 4 \upi R^2(t)\left[\Ppwn(t)-\Pej(R,t)\right].
\label{eqmomentum2}
\end{eqnarray}
In the following we assume, similarly to our former works, that mass accretion onto the shell takes place whenever $\vpwn > \vej$. This is however a delicate issue, and we will devote Sect.~3.4 to discuss more about it.

In previous versions of {\sc TIDE} we did not consider the change from Eq.~\ref{eqvel} to Eq.~\ref{eqvel22} when $dM/dt=0$.
Thus, a negative force --towards the pulsar-- was maintained, as if there was a continuous increase of mass when contraction was ongoing.
This term (the last one in Eq.~\ref{force}) must necessarily overestimate (given that the velocity is negative) the compression of the PWN during reverberation.
For further reference we shall quote the older version of the code as {\sc TIDE} v2.2, while the newer one as {\sc TIDE} v2.3, also dubbed as ``fully conservative momentum treatment''.
In Sect.~3.1 we will compare results using the two different codes.

\subsection{Profiles for the shocked and unshocked medium}
%
This section is devoted to describe how our model quantitatively treats the SNR conditions outside the massive shell.
A required preliminary information is about the position of the forward shock (FS) and the reverse shock (RS), which here are computed following the analytic prescriptions given by \citet{Truelove1999}, as is usual in all PWN literature.

In order to solve the dynamical equations introduced in the previous section, we need  the quantities $\Pej(r,t)$, $\vej(r,t)$ and $\rhoej(r,t)$.
In the unshocked medium, namely inside the RS, these profiles are like those assumed in \citet{Blondin2001}, and read
\begin{eqnarray}
\vej(r,t) & =&r/t\\
\rhoej(r,t) & = & \begin{cases}
A/t^3, & \text{if } r < \vt t
\label{eq:vejunsh} \\
A(\vt/r)^\omega t^{\omega-3}, & \text{if } \vt t < r < \Rrev
\label{eq:rejunsh}
\end{cases} \\
\Pej(r,t) & =&0,
\label{eq:pejunsh}
\end{eqnarray}
while outside the FS the ambient medium is assumed to have a uniform density ($\rhoism$), zero velocity and zero pressure. Here $\omega$ is the SNR envelope density index and
\begin{eqnarray}
A &=&\frac{5(\omega-5)\Esn}{2 \pi \omega\,\vt^5}; \\
\vt &=& \sqrt{\frac{10(\omega-5)\Esn}{3(\omega-3)\Mej}}.
\end{eqnarray}

From an observational perspective, large values of index $\omega$ are preferred, see e.g., \citet{Colgate1969,Chevalier1981b,Chevalier1982,Chevalier1995}, also \citet{Potter2014} and references therein.

For the shocked medium of the SNR (the region between the RS and the FS), we follow the same prescription used in previous similar works, namely profiles for the density, velocity and pressure taken from the Sedov solution \cite{Sedov1959}.
See also \citet{Bandiera1984} for the explicit formulae for a generic ambient density profile ($\rho\propto r^{-s}$), which we have rewritten in Appendix~\ref{sec:a1} for the case of a uniform ambient density ($s=0$).
It must however be clear that during the reverberation phase these asymptotic profiles are far from being reached. 
In fact the Sedov solution strictly holds only in a late, fully relaxed adiabatic phase, and incidentally it applies only to the swept up ambient medium. Instead, as soon as the RS has encountered all the ejecta (or the PWN if there is one) a much more complex transitional phase takes place, in which reflection shocks travel inwards and outwards, before the asymptotic profiles are achieved.
While we are aware that the investigation of such complex phase would require a fully numerical approach (as it will be done in forthcoming papers of this series), in this paper we will use still Sedov profiles (as usually done in all radiative PWN-models literature), or profiles scaled with them, mostly with the aim of testing how robust the results are from changes of those input profiles, and of casting a bridge between past approaches to the problem and our future analyses. 
Moreover the presence of a PWN interacting with the SNR, especially in case of a powerful one, may alter the dynamics of the SNR shell, and the numerical solution extracted from the case of a sole expansion of the SNR might not be a satisfying approximation. When the PWN shell collides with the RS, the \cite{Truelove1999} trajectory is not valid anymore. We shall come back to this issue as a result of the the whole study of our paper series.

\section{Results}
%
We have used our code, as described in the previous section, to derive for a number of cases the evolution of PWN size during the reverberation phase.
In order to summarize the relevance of this evolutionary phase, we will then use a single parameter, the Compression Factor (CF), defined as the ratio between maximum and minimum radius attained by the PWN, during the reverberation phase.
We study the impact of assumptions on this factor next.

In order to compare with our former results \citep{Torres2018b} and discuss our findings, we perform simulations using two sort-of-extreme cases, the Crab Nebula and J1834.9--0846: the former one is the typical example of a PWN fed by a powerful pulsar; while the latter, with a total energetics (estimated as proportional to $L_0 \tau_0$) down by about a factor 50, can be taken as representative of low-energy PWNe.
The parameters that differentiate the two models are those given in Table \ref{tab:pwnpar}, and have been taken from \citet{Torres2018b}.
The containment factor is a global parameter of the fit which defines the maximum ratio between the Larmor radius of the particles and the termination shock radius of the PWN. 
See \citet{Martin2012,Martin2016} for a more detailed definition of all the parameters involved.
In this list, a number of parameters directly affect the extent of synchrotron and inverse Compton radiative losses, and consequently the level of the particles pressure.
Other parameters used with the same values for both models are: SN energy explosion $\Esn=10^{51}\U{erg}$, interstellar medium density $\nism=0.5\U{cm^{-3}}$, SNR envelope density index $\omega=9$, and the PWN and SNR adiabatic indices which are 4/3 and 5/3, respectively.
\begin{table}
\scriptsize
\centering
\caption{Parameters used in the simulations for the Crab Nebula and J1834.9--0846. \label{tab:pwnpar}}
\begin{tabular}{lccc}
\hline
Parameter & Symbol & Crab Nebula & J1834.9--0846\\
\hline
Braking index & $n$ & 2.51 & 2.2\\
Initial spin-down age (yr) & $\tau_0$ & 758 & 280\\
Initial spin-down luminosity (erg s$^{-1}$) & $L_0$ & $3 \times 10^{39}$ & $1.74 \times 10^{38}$\\
SNR ejected mass (M$_\odot$) & $\Mej$ & 9 & 11.3\\
Far infrared temperature (K) & $T\rs{fir}$ & 70 & 25\\
Far infrared energy density (eV cm$^{-3}$) & $w\rs{fir}$ & 0.1 & 0.5\\
Near infrared temperature (K) & $T\rs{nir}$ & 5000 & 3000\\
Near infrared energy density (eV cm$^{-3}$) & $w\rs{nir}$ & 0.3 & 1\\
Energy break & $\gamma_b$ & $9 \cdot 10^{5}$ & $10^{7}$\\
Low energy index & $\alpha_l$ & 1.5 & 1\\
High energy index & $\alpha_h$ & 2.54 & 2.1\\
Containment factor & $\epsilon$ & 0.27 & 0.6\\
Magnetic fraction & $\eta$ & 0.02 & 0.045\\
\hline
\end{tabular}
\end{table}

\subsection{Impact of a fully conservative treatment}

\begin{figure*}
\centering
\includegraphics[width=0.95\textwidth]{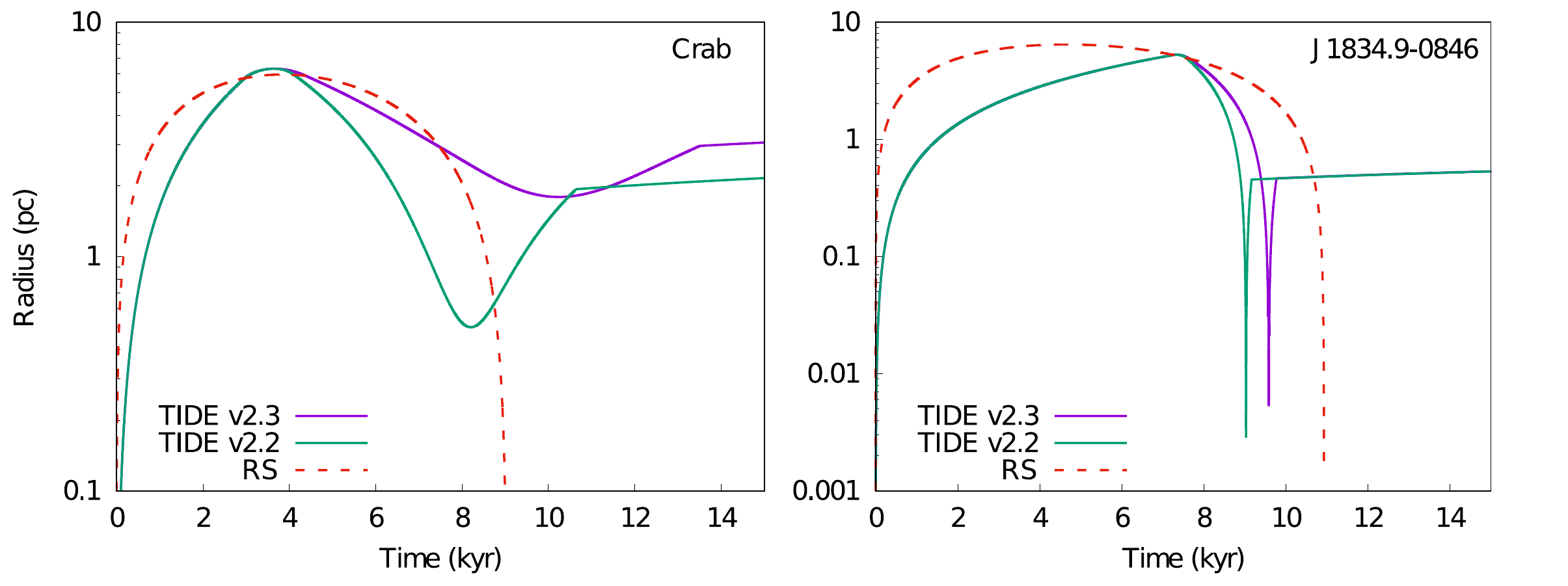}
\caption{Comparison of the radius evolution between {\sc TIDE} v2.3 --fully conservative momentum treatment-- and earlier v2.2, where the momentum equation is not modified according to the evolution stage. Note that Eq.~\ref{eqvel} is implemented in {\sc TIDE} v2.2, while {\sc TIDE} v2.3. adopts Eq.~\ref{eqvel22}. RS stands for the reverse shock. Two representative examples are shown, {\bf Left:} Crab Nebula. {\bf Right:} J1834.9--0846. 
}
\label{newrad}
\end{figure*}

In Sect.~2, we have introduced the fully conservative momentum prescription, which being self-consistent  
is expected to be more accurate by comparison with our former results \citep{Torres2018b}.
We have used it in all forthcoming tests of this work.
But in this section,  in order to understand how large is the impact of this change, and to test the overall validity of our previous analyses, 
we compare our new results with the older next. 
In Figure~\ref{newrad} we show to which extent the introduction of the fully conservative momentum treatment has modified the calculated evolution of the radius up to 15 kyr, for both the case of the Crab Nebula and that of J1834.9--0846.
From the figure, as well as from Table~\ref{tab:tidecomp}), it is apparent that with the fully conservative momentum prescription the compression is always somewhat less efficient.
The largest correction is attained for more powerful PWNe, as it can be seen that the CF for the Crab Nebula is a factor of 3.7 smaller using the new formulation.
On the other hand, in the case of less energetic pulsars the correction is less evident: we may see that the CF for J1834.9--0846 is now only a factor 1.7 smaller.
Therefore, despite the fact that fully conservative momentum always gives smaller CFs, large CFs values are anyway reached for less-energetic pulsars. In particular  there is still a rather wide range of pulsar parameters in which reverberation would ead to periods of superefficiency (when the luminosity in a given frequency band, e.g., X-rays or TeV, exceeds the spin-down power, especially at times close to that of maximum compression, as described in \cite{Torres2018b,Torres2019}).
For instance, already for J1834.9--0846 the maximum luminosity achieved in X-rays exceeds the spin-down power at that time by a factor in excess of a hundred.
For this pulsar in particular, the solution presented in \cite{Torres2017} for describing its magnetar PWN observations (at an age less than 8000 years) is unaffected by changing the treatment to a fully conservative one, given that the PWN current stage is right at the start of the reverberation process, where differences in the evolution are still negligible, see Figure~\ref{newrad}.

\begin{table*}
\small
\centering
\caption{Obtained CFs. \label{tab:tidecomp}}
\begin{tabular}{lcccccc}
\hline
 & \multicolumn{3}{c}{{\sc TIDE} v2.2} & \multicolumn{3}{c}{{\sc TIDE} v2.3} \\
 & $\Rmax$ (pc) & $\Rmin$ (pc) & CF & $\Rmax$ (pc) & $\Rmin$ (pc) & CF \\
\hline
Crab Nebula &  6.326 &  0.499 &  12.68 &  6.329 &  1.793 &  3.530 \\
J1834.9--0846 & f 5.270 &  0.003 &  1757 &  5.270 &  0.005 &  1054 \\
\hline
\end{tabular}
\end{table*}

\subsection{Impact of the original ejecta density profile}
\begin{figure*}
\centering
\includegraphics[width=0.95\textwidth]{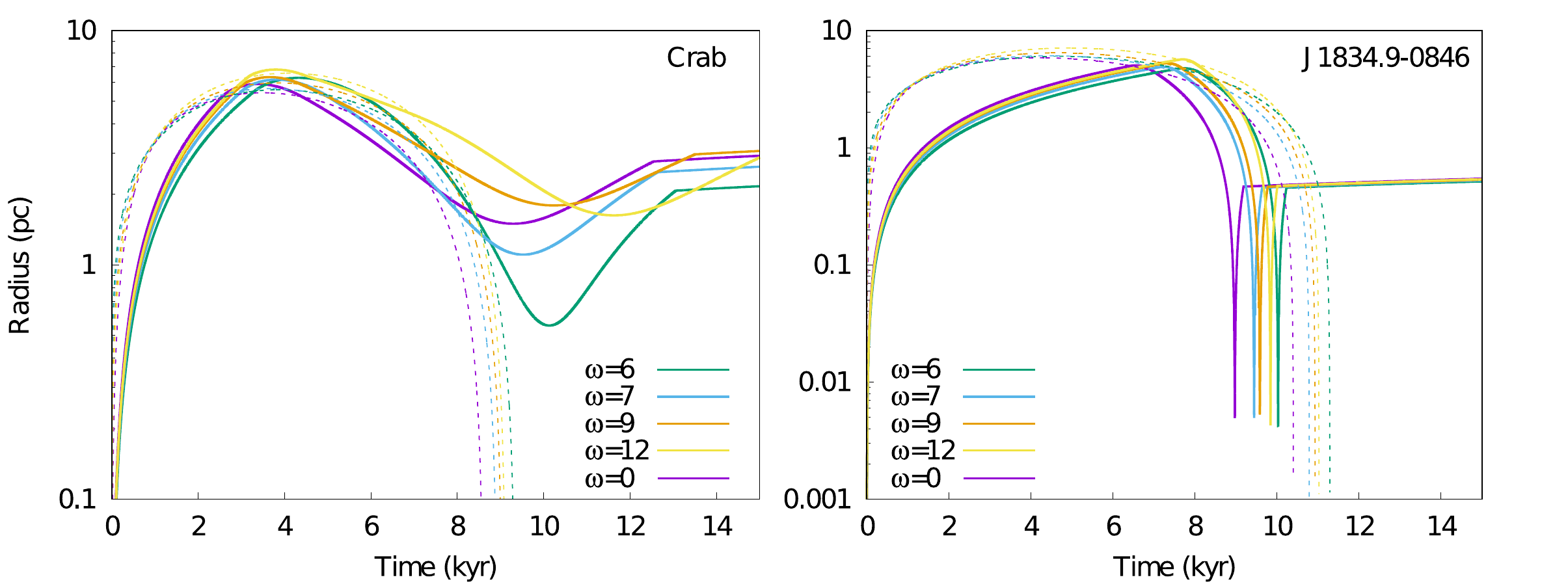}
\caption{Comparison of the radius evolution as a function of the SNR envelope density index. {\bf Left:} Crab Nebula. {\bf Right:} J1834.9--0846. The reverse shock position is shown in dashed lines in each case.}
\label{fig:index}
\end{figure*}

%
Figure~\ref{fig:index} shows the evolution in size, as from our simulations, for the Crab Nebula and J1834.9--0846, with $\omega$ ranging from 6 to higher values (the case $\omega = 0$ can be seen as the limit for $\omega \rightarrow \infty$).
The dashed lines in the figure represent the trajectory of the RS following the prescription given by \citet{Truelove1999}.
Note that in this figure, as well a in some of the following ones, there are times in the reverberation phase in which the PWN size is smaller than the RS one. This should not be surprising, because the \citet{Truelove1999} solutions assume a non-radiative SNR with no PWN inside. In the presence of a PWN the RS has a meaning only before the beginning of the reverberation (see further discussion above).

The results on the compression factor, extracted from these curves, are summarized in Table~\ref{tab:index}.
For both the Crab Nebula and J1834.9--0846 models, the PWN is maximally compressed for $\omega=6$, while the minimally compressed cases are found for larger values of the SNR density index ($\omega =9$ and $\omega=7$, respectively), before increasing around $\omega=12$, and eventually decreasing again at the asymptotic value $\omega=0$.

It may be noticed that the evolution with $\omega$ of the RS and PWN curves does not appear to be monotonic , but such behavior is not fully understandable.
For instance, one should have expected the line for $\omega=12$ to be the closest to that for $\omega=0$, but from the figure it is evident this is not the case.
The issue whether it derives from a real physical effect, or not rather from low accuracy of the input formulae for the RS evolution, will be discussed in a future paper of the series.

\begin{table}
\small
\centering
\caption{Obtained CFs for different values of the SNR density index $\omega$. \label{tab:index}}
\begin{tabular}{lcccccc}
\hline
 & $w=0$ & $w=6$ & $w=7$ & $w=9$ & $w=12$ \\
\hline
{\bf Crab Nebula} & & & & & \\
$\Rmax$ (pc) &  5.912 &  6.281 &  6.177 &  6.329 &  6.816\\
$\Rmin$ (pc) &  1.499 &  0.551 &  1.106 &  1.793 & 1.626\\
CF & 3.944 &  11.40 &  5.585 &  3.530 &  4.192\\
{\bf J1834.9--0846} & & & & & \\
$\Rmax$ (pc) &  5.041 & 4.765 &  4.902 & 5.270 & 5.676\\
$\Rmin$ (pc) & 0.005 &  0.004 &  0.005 & 0.005 & 0.004\\
CF & 1008 & 1191 & 980.4 & 1054 & 1419 \\
\hline
\end{tabular}
\end{table}

\subsection{Impact of the shocked medium conditions}

\begin{figure*}
\centering
\includegraphics[width=0.95\textwidth]{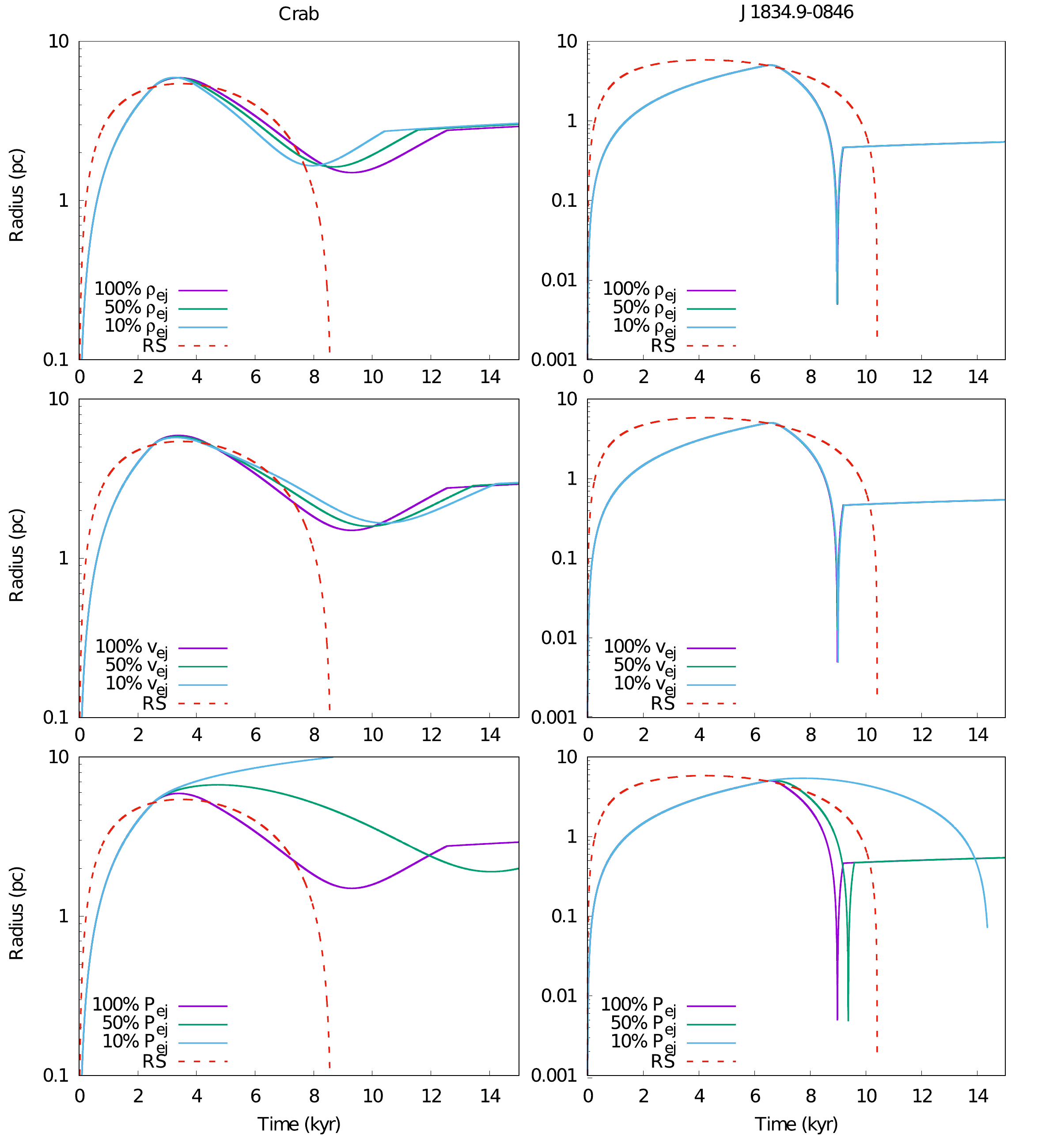}
\caption{Radius evolution varying the $\rhoej$, $\vej$ and $\Pej$ profiles
for the Crab Nebula (left) and J1834.9--0846 (right). The dashed red line represents the reverse shock of the SNR (RS).}
\label{fig:prof_var}
\end{figure*}

\begin{figure*}
\centering
\includegraphics[width=0.95\textwidth]{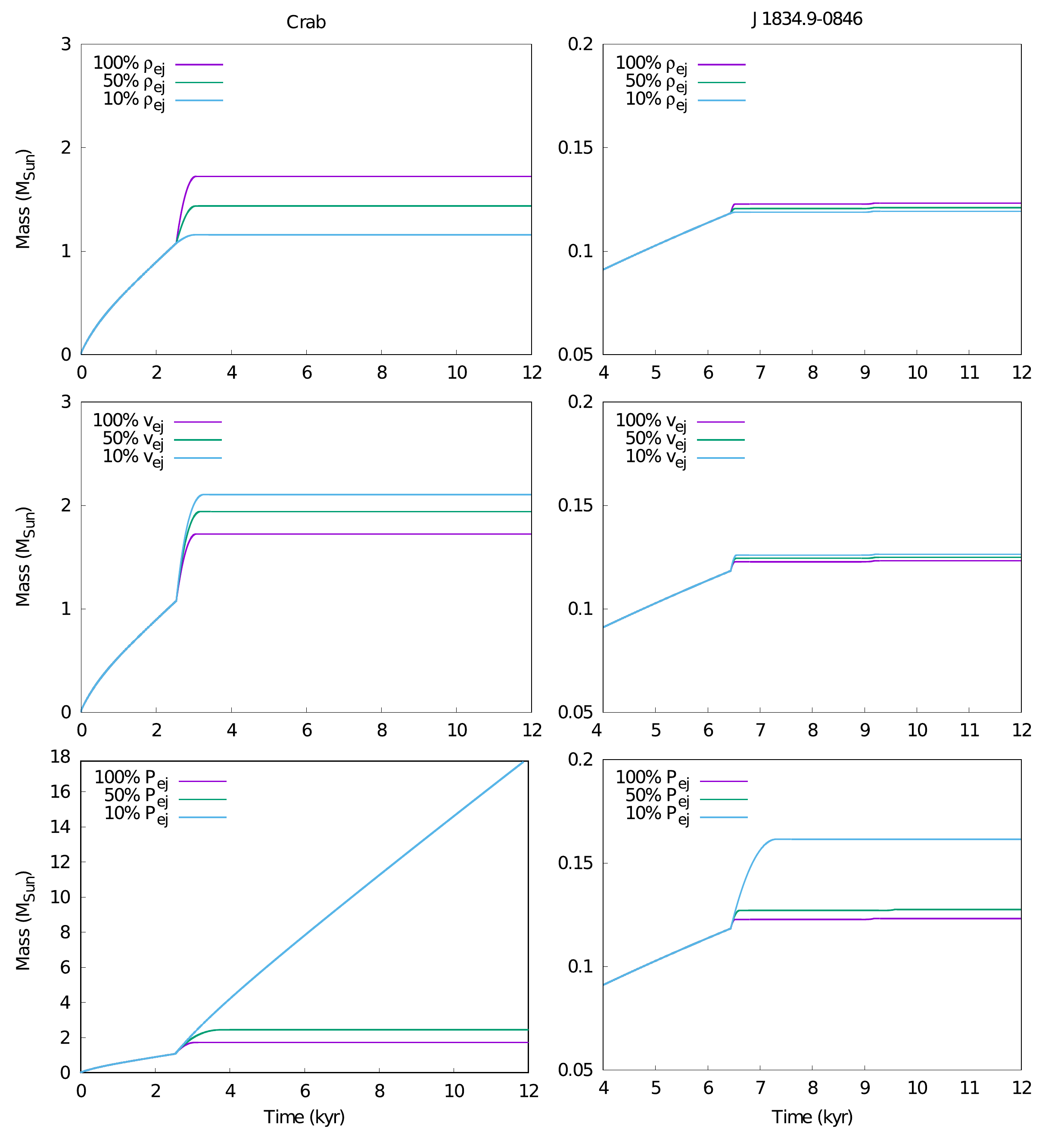}
\caption{Mass shell evolution varying the $\rhoej$, $\vej$ and $\Pej$ profiles for the Crab Nebula (left) and J1834.9--0846 (right). A dramatic reduction of the ejecta pressure, down to the 10\% of the original value, delays the beginning of the reverberation or, as in the case of the Crab (bottom left panel) it can be even cancelled at all.}
\label{fig:prof_var_mass}
\end{figure*}

As already mentioned above, a crucial issue of the present approach resides in the fact that the profiles of density, velocity and pressure in the shocked SNR material are poorly modelled.
Here we will tackle the problem in a different way, namely by testing how sensitive the CF is to changes in the profiles of the various quantities.
We artificially modify such profiles by shifting their levels to respectively 50\% and 10\% of their original values.
We first vary just one input physical quantity at a time, either $\rhoej$, $\vej$ or $\Pej$.
A similar prescription was discussed in \citet{Bucciantini2011}, where the pressure in the ejecta was reduced by a factor of 50\% with respect to the Sedov one, using as a reference the pressure extrapolated form a 1D numerical model in \citep{Bucciantini:2003}.
In Figures~\ref{fig:prof_var} and \ref{fig:prof_var_mass} we show the evolution of the radius and mass shell for each case;  while Table \ref{tab:band_prof} summarises the results.

Let us begin with a qualitative discussion of these figures.
It is apparent from Figure~\ref{fig:prof_var} that the evolution of the PWN during the reverberation is mostly sensitive to changes of the pressure in the ejecta immediately outside the massive shell, and much less to those of density and velocity.
The qualitative effect of a change of the pressure is rather obvious: with a lower outer pressure the deceleration effect on the massive shell is less effective, so that the compression may be delayed, of a lower amount, and even disappear at all (see the model of the Crab Nebula with only 10\% of the standard $\Pej$).

The  changes in $\rhoej$ and $\vej$ have a more direct effect in ruling the evolution of the mass collected into the shell. Reaching a higher mass implies a lower efficiency in the deceleration, keeping the outer pressure constant.
This easily explains why the compression starts earlier for lower $\rhoej$ values, and instead starts later for lower $\vej$ values (one should keep in mind that a lower $\vej$ implies a higher relative velocity of the shell with respect to the ejecta).
It is also clear why the effect of $\rhoej$ and $\vej$ changes are almost negligible for the case of J1834.9--0846.
Figure~\ref{fig:prof_var_mass} shows that the increase of shell mass during the reverberation phase is small, compared to the case of the Crab Nebula.
So, in the case of J1834.9--0846 the shell reacts in the same way to the outer pressure, almost independently of the values taken for $\rhoej$ and $\vej$.

Finally, the result that the mass accretion is lower in the case of a higher outer pressure is simply due to the fact that a higher pressure implies a higher deceleration of the shell, and therefore its relative velocity vanishes more rapidly and further mass accretion is quenched.
Incidentally, Figure~\ref{fig:prof_var_mass} also shows the plateau in the shell mass evolution due to the fact that further mass accretion is inhibited when $\vpwn$ becomes smaller than $\vej$.
%
%
It is interesting to note (see Table \ref{tab:band_prof}) that only the values for the maximum and minimum radii may change significantly by changing the model parameters, while the CFs are only mildly sensitive to changes in the profiles.

\begin{table*}
\small
\centering
\caption{Obtained CFs changing the values of the \citet{Bandiera1984} profiles for the shocked medium density, velocity and pressure. \label{tab:band_prof}}
\begin{tabular}{lccccccc}
\hline
 & Normal & $50\%\ \rhoej$ & $10\%\ \rhoej$ & $50\%\ \vej$ & $10\%\ \vej$ & $50\%\ \Pej$ & $10\%\ \Pej$ \\
\hline
{\bf Crab Nebula ($\omega=0$)} & & & & & & & \\
$\Rmax$ (pc) & { 5.912} & {5.919} & { 5.931} & { 5.822} & {5.737} & {6.697} & -* \\
$\Rmin$ (pc) & { 1.499} & { 1.628} & { 1.656} & { 1.590} & { 1.666} & { 1.906} & -* \\
CF & { 3.944} & { 3.636} & { 3.581} & { 3.662} & { 3.444} & { 3.514} & -* \\
{\bf J1834.9--0846 ($\omega=0$)} & & & & & & & \\
$\Rmax$ (pc) & { 5.041} & { 5.041} & { 5.041} & { 5.040} & { 5.040} & { 5.086} & { 5.430**} \\
$\Rmin$ (pc) & { 0.005} & { 0.005} & { 0.005} & { 0.005} & { 0.005} & { 0.005} & -** \\
CF & { 1008} & { 1008} & { 1008} & { 1008} & { 1008} & { 1017} & -** \\
{\bf Crab Nebula ($\omega=9$)} & & & & & & & \\
$\Rmax$ (pc) & {6.329} & { 6.336} & { 6.350} & { 6.247} & { 6.173} & {6.924} & -* \\
$\Rmin$ (pc) & { 1.793} & { 1.961} & { 1.984} & { 1.765} & { 1.640} & { 1.555} & -* \\
CF & { 3.530} & {3.231} & { 3.201} & { 3.539} & { 3.742} & { 4.453} & -* \\
{\bf J1834.9--0846 ($\omega=9$)} & & & & & & & \\
$\Rmax$ (pc) & { 5.270} & { 5.270} & {5.270} & { 5.269} & {5.269} & { 5.299} & { 5.523**} \\
$\Rmin$ (pc) & {0.005} & { 0.005} & { 0.005} & { 0.005} & { 0.005} & { 0.005} & -** \\
CF & {1054} & {1054} & {1054} & {1054} & { 1054} & { 1060} & -** \\
\hline
\multicolumn{8}{l}{* Reverberation phase starts later than 45 kyr}\\
\multicolumn{8}{l}{** The shocked profile is not defined for radii so close to the center of the SNR.}
\end{tabular}
\end{table*}

\subsection{Impact of the mass of the ejecta and density of the ISM}

In the absence of a PWN, the effects of the mass of the ejecta and density of the ISM are quite easy to understand.
Once  a characteristic time scale and a characteristic length scale are set, on the basis of the dimensional parameters of the problem, the dimensionless solution for the SNR, including the trajectories of all shocks, is unique (for a given value of $\omega$).
In particular, following \citet{Truelove1999}, the characteristic radius and time are given by
\begin{eqnarray}
	\Rch &=& \Mej^{1/3}\rhoism^{-1/3}\,;	\\
	\tch &=& \Mej^{5/6}\Esn^{-1/2}\rhoism^{-1/3}\,.
\end{eqnarray}
In the presence of a PWN the situation is more complex to describe: further dimensional parameters are introduced by the pulsar, so that the problem is no longer self-similar.

Let us focus on J1834.9--0846, as an example of less powerful pulsars, and let us make the mass of the ejecta and density of the ISM  changing  both of 50\% and 150\% with respect their original values.
The results are displayed in Table \ref{inf-ism}.
From a power-law fit of these results we find $\Rmax\propto\Mej^{0.343}\rhoism^{-0.335}$, namely with a scaling very close to that for $\Rch$.
This correspondence can be justified in the following way: for an object like J1834.9--0846, the PWN-RS interaction begins after the RS has already started contracting, but has not contracted too much, and the PWN is not powerful enough to force a re-expansion of the interface.
Therefore approximating $\Rmax$ as a a given fraction of $\Rch$ in this case is an excellent approximation, as it can be seen from the last row of Table \ref{inf-ism}.

The case for $\Rmin$ variations is more difficult to interpret, because there are more variables to take into account (i.e., the pressure of the ejecta, the adiabatic energy gain and the mass of the PWN shell), but in the case of J1834.9--0846 making that slight variations on it affects critically to the CF obtained.

\begin{table*}
\small
\centering
\caption{Obtained CFs changing the values of the ejecta mass and ISM density for J1834.9--0846 ($\omega=9$). \label{tab:j1834}}
\begin{tabular}{lccccc}
\hline
 & Normal & $50\%\ \rhoism$ & $150\%\ \rhoism$ & $50\%\ \Mej$ & $150\%\ \Mej$ \\
\hline
$\Rmax$ (pc) & 5.270 & 6.386 & 4.387 & 3.991 & 5.769 \\
$\Rmin$ (pc) & 0.005 & 0.005 & 0.005 & 0.006 & 0.005 \\
CF & 1054 & 1277 & 877.4 & 665.2 & 1154 \\
$\Rch$ (pc) & 9.730 & 12.26 & 8.500 & 7.723 & 11.14 \\
$\Rmax/\Rch$ & 0.5416 & 0.5209 & 0.5161 & 0.5168 & 0.5179 \\
\hline
\label{inf-ism} 
\end{tabular}
\end{table*}

\subsection{Testing prescriptions for the shell mass evolution}
%
A non-trivial issue, within this approach, is to `decide' when the mass accretion starts and ends.
Both self-similar models  and numerical simulations (see e.g., \citealt{Jun1998}) show that, as long as unshocked ejecta are swept up by the PWN, the mass shell that is formed is truly a thin shell, with a relative width smaller than a few percent.
This is due to a combination of factors, the leading one being the fast decrease (like $t^{-3}$; see next section) of the density of the unshocked ejecta.
This, however, will no longer be true after the shell has been reached by the the RS, and starts interacting with the shocked ejecta: after that time, rather than a further accretion of SNR matter onto the shell, a reflected shock appears, propagating outwards into the thermal SNR the information about the impact against that massive shell.
As a consequence, 0+1 models are intrinsically caveated to provide an accurate description of the dynamical part.
We shall study this in detail in future papers of the series, where we shall interface {\sc TIDE} with a 1D hydrodynamic numerical code.

So far, 0+1 dynamical-code based approaches to this problem, by our group as well as by many others who confronted it earlier, see e.g., \cite{Gelfand2009}, have been complemented with the prescription that mass accretion stops at all times at which $\vpwn(t) < \vej(R(t),t)$, thus assuming that accretion always leads to a thin shell, even if it is not true.
Alternately, one can assume that the mass accretion onto the shell stops at the beginning of the reverberation phase, namely when we start having $\Rrev \le \Rpwn$ \citep{Bucciantini:2011}.
In this case the thin-shell approximation is more closely satisfied, but the medium immediately outside the shell is no longer correctly described by using the pristine SNR profile.
In this section we compare the results obtained with these two different prescriptions, and taking their difference as a reasonable estimate of the uncertainties involved.

\begin{figure*}
\centering
\includegraphics[width=0.95\textwidth]{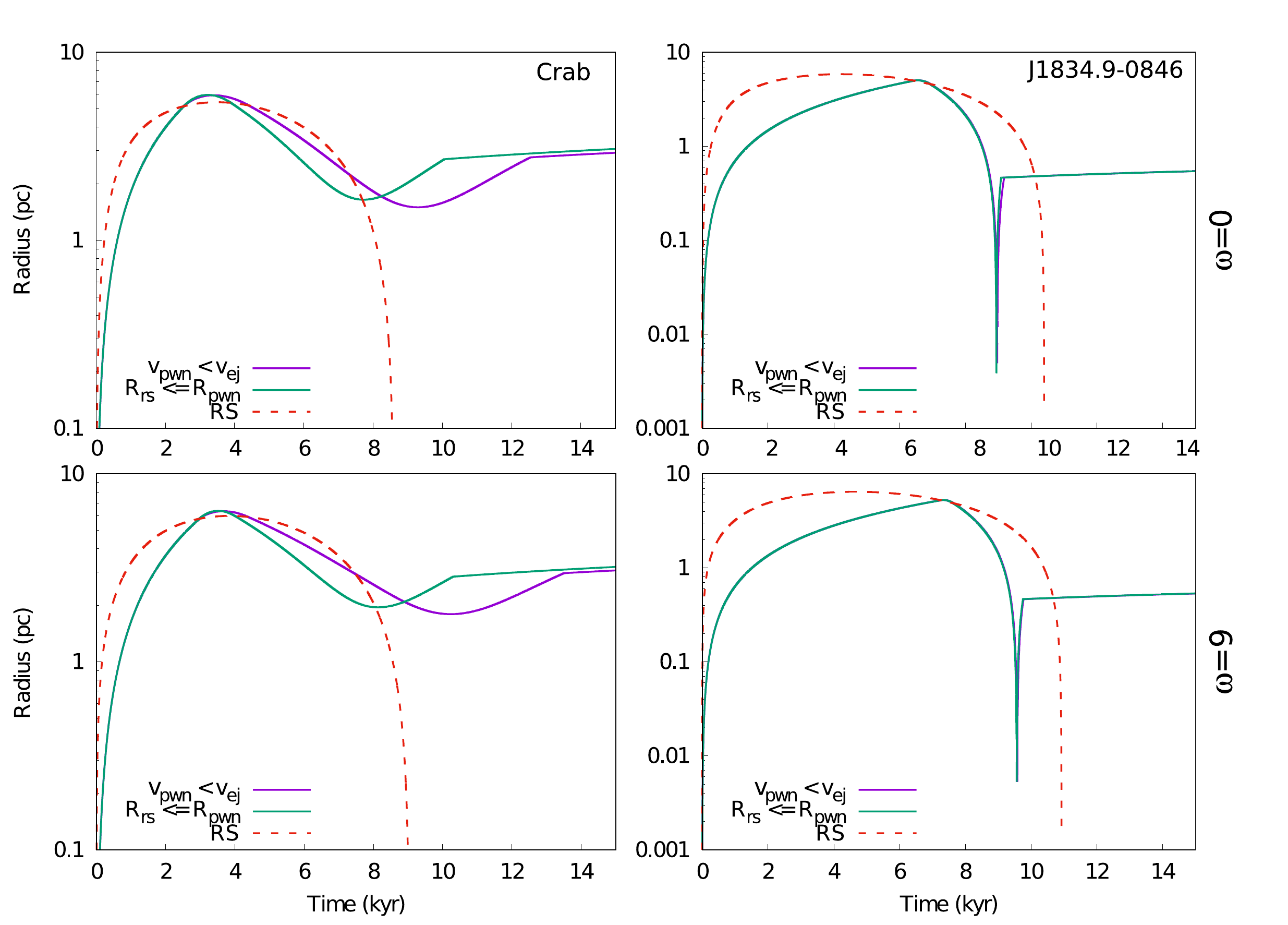}
\caption{Evolution of the PWN radius for the Crab Nebula (top) and J1834.9--0846 (bottom) if we apply $dM/dt=0$ when $\vpwn < \vej$ and $\Rrev \le \Rpwn$, respectively. Left-hand plots correspond to the case of the Crab nebula, right-hand ones to the case of J1834.9--0846. The top row is for a SNR density profile index $\omega=0$, bottom row for $\omega=9$.}
\label{testrrs}
\end{figure*}

Figure~\ref{testrrs} shows that imposing $dM/dt=0$ with $\Rrev < \Rpwn$ leads to a slightly earlier compression of the PWN, essentially because stopping the mass accretion at an earlier time results in a lower inertia for the massive shell. Such effect is much more prominent in the Crab with respect to J1834.9--0846, since in the latter object the mass accretion after the beginning of reverberation is small even with our original prescription.
Table \ref{tab:r=rs}) shows that the two different prescriptions for the mass accretion give anyway very similar CF values.
This is not a formal proof that the resulting CF values are rather insensitive to the prescription used, but make us feel comfortable that the CF values obtained are rather stable and reliable.
In fact, for J1834.9--0846 it is actually the same CF, what can be explained by a minimal change found in the mass loading of the shell in this case.

\begin{table}
\small
\centering
\caption{Obtained CFs applying $dM/dt=0$ when $\vej < \vpwn$ (original approach) and $\Rrev \le \Rpwn$.
\label{tab:r=rs}}
\begin{tabular}{lcc}
\hline
 & $\vej < \vpwn$ & $\Rrev \le \Rpwn$ \\
\hline
{\bf Crab Nebula} & & \\
$\Rmax$ (pc) & { 6.329} & { 6.356} \\
$\Rmin$ (pc) & { 1.793} & { 1.949} \\
CF & { 3.530} & { 3.261} \\
{\bf J1834.9--0846} & & \\
$\Rmax$ (pc) & { 5.270} & { 5.270} \\
$\Rmin$ (pc) & { 0.005} & { 0.005} \\
CF & { 1054} & { 1054} \\
\hline
\end{tabular}
\end{table}

\subsection{How much compression is radiation-produced?}
%
Given that the existence of radiation enables a channel for energy loss, we studied how much of the compression attained in each case is radiation-produced.
In practice, we consider the same radiative simulations but assuming a null magnetic fraction $\eta$ (so that there is no magnetic field) and extremely low energy densities for the background photons as well, so that there are no inverse Compton losses.
We are also avoiding here the existence of escaping particles.
As a further check of the dynamical model, we consider Chevalier's set of equations \citep{Chevalier2005}, which consists of Eqs.~\ref{eqmass}, \ref{eqvel}, with the addition of
\begin{equation}
\frac{d}{dt}(4 \upi R^3(t)\Ppwn(t))=L(t)-4 \upi R^2(t)\Ppwn(t)\frac{dR(t)}{dt}
\label{eq:energy}
\end{equation}
for modelling the evolution of the total PWN pressure in the absence of radiative losses: this last equation is equivalent to Eq.~\ref{eqPBevol}, used for the magnetic pressure.
In order to solve this set of equations, which is devoid of any radiative content from the start (a fully non-radiative PWN) we use a separate code, different from TIDE, which does not require the explicit calculation of the evolution of the particles energy distribution.
In this way we have devised a simple benchmark to test, in the non-radiative limit, the numerical accuracy of TIDE.

\begin{figure*}
\centering
\includegraphics[width=0.95\textwidth]{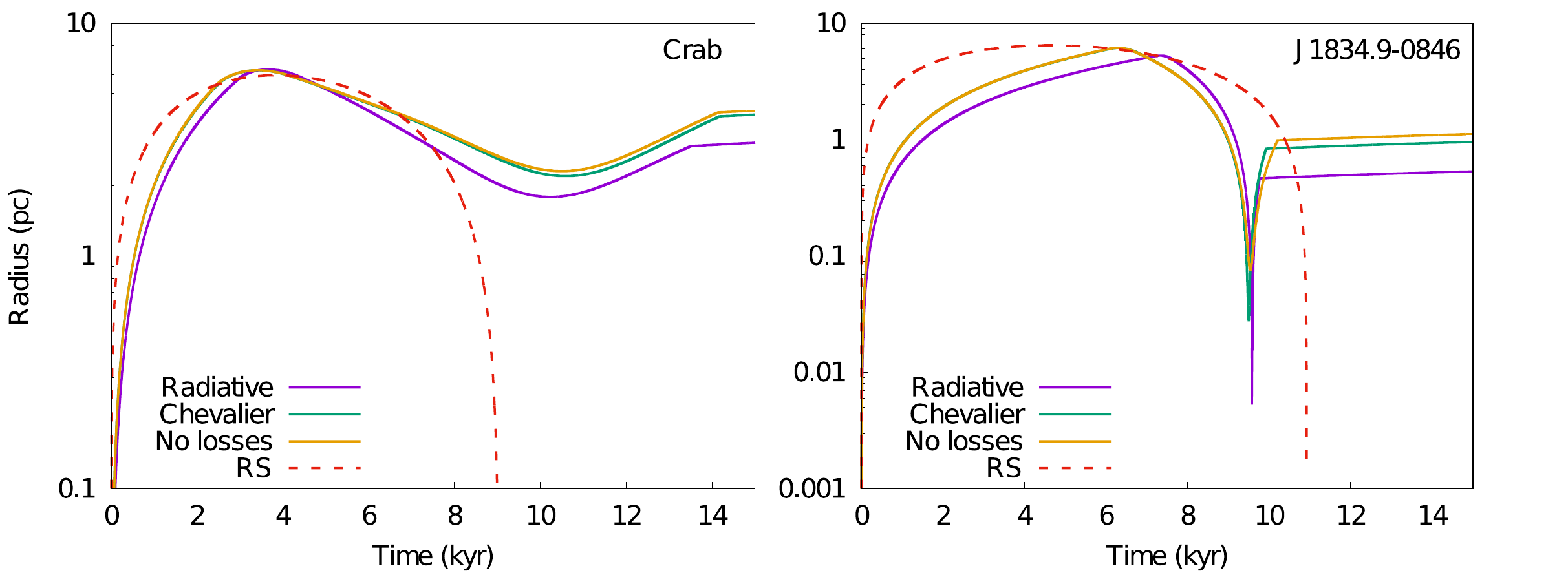}
\caption{Evolution of the radius for radiative and non-radiative models (Chevalier's model representing a benchmark for for the latter ones), for the Crab Nebula (left) and J1834.9--0846 (right). These calculations are performed for a $\omega=9$ envelope density profile.}
\label{test_radius}
\end{figure*}

A comparison of the evolution of $\Rpwn(t)$ computed with the different models (radiative TIDE, TIDE without losses, and Chevalier's respectively) is shown in Figure~\ref{test_radius}).
First, the close resemblance between the Chevalier's code and TIDE without losses represents a good test of the numerical accuracy of our code.
TIDE solves Chevalier's equations not using equation~\ref{eq:energy}, but integrating the particle distribution function in energy.
This approach allows to take into account radiative losses in the calculation of the PWN internal energy.
This difference in the numerical approach introduces a small mismatch between the two codes, specially when reverberation starts, where the integration of the particle diffusion-loss equation is less accurate due to the fast increase of the adiabatic energy gain.

The radiative models may produce significantly more compression than the non-radiative ones (what is to be expected on physical grounds, of course) and that this difference depends on the spin-down power, being larger for less powerful pulsars.
This is shown in Table \ref{CF-rad-norad} for the two examples chosen here (other PWNe we have studied behave similarly).

We also note that some differences between the radiative and non-radiative cases appear before the beginning of reverberation.
This is due to the fact that at very early times, when the PWN magnetic field is large, radiative losses are always important.
How relevant they will be on the overall evolution mostly depends on the values of $L_0$, $\tau_0$, $\eta$, and $\gamma_b$, the energy at which most of the energy of the particle spectrum is.
Figure~\ref{test_radius} shows that, even if with a lower $L_0$, the early--time radiative correction for J1834.9--0846 is larger than that for the Crab.
The reason of this is that the best--fit model for J1834.9--0846 has also a much smaller value for $\tau_0$, and a much larger value for $\gamma_b$ (see Table~\ref{tab:pwnpar}), and both parameters  contribute substantially to increment the importance of radiative losses.
As a result, the effective energy deposited into the shell is lower and consequently, as shown in Figure~\ref{test_radius}, its radius at a given time is also lower.
\begin{table}
\small
\centering
\caption{Obtained CFs for each model ($\omega=9$).}
 \label{CF-rad-norad}
\begin{tabular}{lccc}
\hline
 & Radiative & Chevalier & No losses \\
\hline
{\bf Crab Nebula} & & & \\
$\Rmax$ (pc) & { 6.329} & { 6.274} & { 6.274} \\
$\Rmin$ (pc) & { 1.793} & { 2.204} & { 2.313} \\
CF & { 3.530} & { 2.847} & { 2.712} \\
{\bf J1834.9--0846} & & & \\
$\Rmax$ (pc) & { 5.270} & { 6.121} & { 6.137} \\
$\Rmin$ (pc) & { 0.005} & { 0.028} & { 0.076} \\
CF & { 1054} & { 218.6} & { 80.75} \\
\hline
\end{tabular}
\end{table}

\begin{figure*} 
\centering
\includegraphics[width=0.95\textwidth]{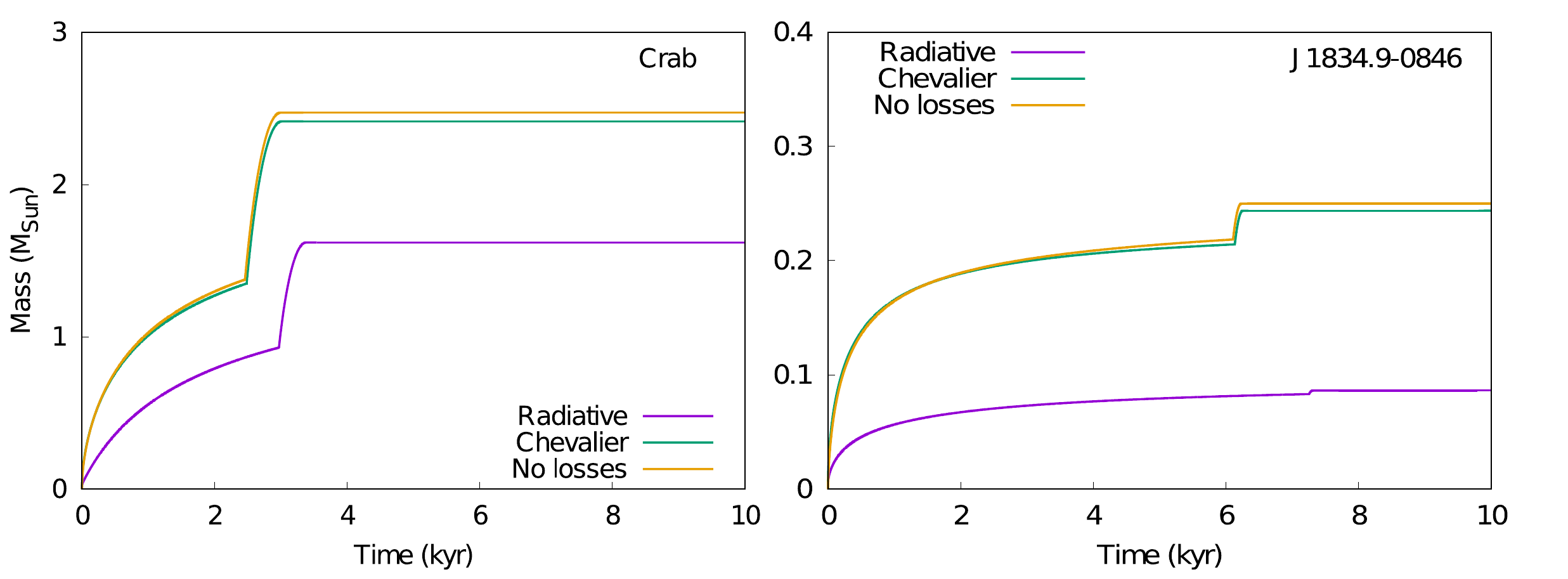}
\caption{Evolution of the mass of the shell for the Crab Nebula (left) and J1834.9--0846 (right). These calculations are performed for a $\omega=9$ envelope density profile.}
\label{test_mass}
\end{figure*}

Figure~\ref{test_mass} shows a consistent increase of mass in the PWN shell as well, for the non-radiative cases compared to the radiative ones.
Before the beginning of reverberation, the higher radial expansion is amplified by the fact that the swept-up mass is proportional to $R^3$; as a direct consequence of this, reverberation also begins earlier.
In our simulations the further mass accretion, taking place during the first part of the reverberation, is also larger in the non-radiative cases: this is justified by the fact that the higher-mass shell decelerates less efficiently, and therefore it continues collecting material at a higher rate, before the compression starts taking place.
The losses via radiation play a dominant role also during the phase of maximum compression, then making the PWN reaction weaker and the CF larger in the radiative case.

\section{Conclusions}

Let us summarize our main conclusions:

\begin{enumerate}
\item A correct treatment of the conservation of the momentum 
reduces the CF (see Table \ref{tab:tidecomp} and Figure~\ref{newrad}), 
especially for some very high spin-down luminosity pulsars like Crab. In low spin-down luminosity ones, the variation is not so important, although still relevant. 

\item This reduction in the compression provided by the former consideration does not eliminate the superefficiency behavior earlier found by \cite{Torres2018b}, being the CFs nevertheless very high. The large CFs (and superefficiency) shown up in advanced radiative models such as those of \cite{Gelfand2009,Martin2016} thus remain as a feature of radiative 0+1 approach to study PWN.
Assessing it further requires a number of significant advances in PWN evolution modelling, understanding better the position and size of the shock regions and how mass loaded they are.

\item  The evolution of the PWN is very sensitive to the ejecta pressure profile, being, on the contrary, only poorly affected by variations of the normalization of the ejecta velocity and density profiles. These variations modify the way the mass is collected into the shell at the PWN boundary, and this affects the deceleration efficiency during the reverberation phase. However, this effect scales with the fraction of mass collected in this late phase, and may then be negligible for less powerful pulsars.

\item CFs are mildly sensitive to reasonable changes in the pressure, density and velocity of the SNR shocked medium (see Table \ref{tab:band_prof} and Figure~\ref{fig:prof_var}).
In fact, in these cases, only the position of the maximum and minimum radii change significantly, especially when the ejecta pressure is changed. 
 
\item The same happens with the mass of the PWN shell due to the ability to go deeper into the ejecta when the pressure is lower, or the contrary when it is higher (see Figure~\ref{fig:prof_var_mass}).

\item As expected, there is a rather regular dependence of the CF with the variation of the ejected mass and the ISM density (see Table \ref{inf-ism}).

\item Changing the criterion for accreting SNR material to the start of the reverberation (when $\Rrev \le \Rpwn$), instead of $\vej < \vpwn$ leads to no appreciable change in the CFs (see Table \ref{tab:r=rs} and Figures \ref{testrrs}).

\item Considering the same PWN with and without radiation (although of course, radiation is there in the physical world), the former compresses much more due to the extra losses (see Table \ref{CF-rad-norad} and Figure~\ref{test_radius}). This is something that has to be taken into account when comparing with hydrodynamical models (devoid of radiative content), as we shall discuss in forthcoming papers of the series. This difference can be large (a factor of 10 or more) for particular PWN cases with low spin-down power. 
 
\item In order to get 0+1 models compatible with the results obtained by 1D or higher-order hydrodynamical simulations, it will be crucial to find a good representation of the pressure of the ejecta too.
After the RS and the PWN collide, radiative models assuming their existence as separate entities are necessarily inaccurate. A detailed comparison between hydrodynamical simulations and radiative models are needed in order to calibrate the level of approximation done by the latter assumption and prescribe a new way (if possible) to deal with this effect in radiative models. 
\end{enumerate}
This analysis shows that, while using one-zone models in the characterization of the detections of gamma-ray instruments and/or their surveys; and of course their simulated data caution should be payed  in the underlying choices that go into building those models and in the general bias they might introduce in their results.
In particular the assumption of the bounding SNR to be in a the relaxed Sedov state must be handled with care. 
The dynamics of the swept-up shell appears in fact to be very sensitive to the ejecta profiles, in particular to those of the pressure. 
Important variations of the profiles may introduce an extended duration of the reverberation, changing the physics of the system -- in particular when considering radiation -- even if the final CF results to be only mildly sensitive to those changes.
We will discuss a more appropriate description of the SNR properties in a forthcoming paper of the same series.

\section*{Acknowledgements}

This work has been supported by grants PGC2018-095512-B-I00, SGR2017-1383, AYA2017-92402-EXP, 2017-14-H.O ASI-INAF and INAF MAINSTREAM.


\bibliographystyle{mn2e}
\bibliography{pwn}

\begin{thebibliography}{}

\bibitem[\protect\citeauthoryear{{Abdalla}, {Abramowski} A.~{Aharonian} \& {et
  al. for the H.E.S.S. collaboration}}{{Abdalla}
  et~al.}{2018}]{H.E.S.S.-Collaboration:2018a}
{Abdalla} H.,  {Abramowski} A.~{Aharonian} F.,    {et al. for the H.E.S.S.
  collaboration} 2018, \aap, 612, A2

\bibitem[\protect\citeauthoryear{{Bandiera}}{{Bandiera}}{1984}]{Bandiera1984}
{Bandiera} R.,  1984, \aap, 139, 368

\bibitem[\protect\citeauthoryear{{Barkov} \& {Lyutikov}}{{Barkov} \&
  {Lyutikov}}{2018}]{Barkov2018}
{Barkov} M.~V.,  {Lyutikov} M.,  2018, ArXiv e-prints

\bibitem[\protect\citeauthoryear{{Bednarek} \& {Bartosik}}{{Bednarek} \&
  {Bartosik}}{2003}]{Bednarek:2003}
{Bednarek} W.,  {Bartosik} M.,  2003, \aap, 405, 689

\bibitem[\protect\citeauthoryear{{Blondin}, {Chevalier} \&
  {Frierson}}{{Blondin} et~al.}{2001a}]{Blondin:2001}
{Blondin} J.~M.,  {Chevalier} R.~A.,    {Frierson} D.~M.,  2001a, \apj, 563,
  806

\bibitem[\protect\citeauthoryear{{Blondin}, {Chevalier} \&
  {Frierson}}{{Blondin} et~al.}{2001b}]{Blondin2001}
{Blondin} J.~M.,  {Chevalier} R.~A.,    {Frierson} D.~M.,  2001b, \apj, 563,
  806

\bibitem[\protect\citeauthoryear{{Bucciantini}}{{Bucciantini}}{2013}]{Bucciantini:2013a}
{Bucciantini} N.,  2013, in {Ness} J.~U.,  ed., The Fast and the Furious:
  Energetic Phenomena in Isolated Neutron Stars, Pulsar Wind Nebulae and
  Supernova Remnants {Review of the Theory of PWNe}.
p.~2

\bibitem[\protect\citeauthoryear{{Bucciantini}, {Arons} \&
  {Amato}}{{Bucciantini} et~al.}{2011a}]{Bucciantini2011}
{Bucciantini} N.,  {Arons} J.,    {Amato} E.,  2011a, \mnras, 410, 381

\bibitem[\protect\citeauthoryear{{Bucciantini}, {Arons} \&
  {Amato}}{{Bucciantini} et~al.}{2011b}]{Bucciantini:2011}
{Bucciantini} N.,  {Arons} J.,    {Amato} E.,  2011b, \mnras, 410, 381

\bibitem[\protect\citeauthoryear{{Bucciantini}, {Bandiera}, {Blondin}, {Amato}
  \& {Del Zanna}}{{Bucciantini} et~al.}{2004}]{Bucciantini:2004}
{Bucciantini} N.,  {Bandiera} R.,  {Blondin} J.~M.,  {Amato} E.,    {Del Zanna}
  L.,  2004, \aap, 422, 609

\bibitem[\protect\citeauthoryear{{Bucciantini}, {Blondin}, {Del Zanna} \&
  {Amato}}{{Bucciantini} et~al.}{2003}]{Bucciantini:2003}
{Bucciantini} N.,  {Blondin} J.~M.,  {Del Zanna} L.,    {Amato} E.,  2003,
  \aap, 405, 617

\bibitem[\protect\citeauthoryear{{Chevalier}}{{Chevalier}}{1981}]{Chevalier1981b}
{Chevalier} R.~A.,  1981, \apj, 246, 267

\bibitem[\protect\citeauthoryear{{Chevalier}}{{Chevalier}}{1982}]{Chevalier1982}
{Chevalier} R.~A.,  1982, \apj, 258, 790

\bibitem[\protect\citeauthoryear{{Chevalier}}{{Chevalier}}{2005}]{Chevalier2005}
{Chevalier} R.~A.,  2005, \apj, 619, 839

\bibitem[\protect\citeauthoryear{{Chevalier} \& {Dwarkadas}}{{Chevalier} \&
  {Dwarkadas}}{1995}]{Chevalier1995}
{Chevalier} R.~A.,  {Dwarkadas} V.~V.,  1995, \apjl, 452, L45

\bibitem[\protect\citeauthoryear{{Colgate} \& {McKee}}{{Colgate} \&
  {McKee}}{1969}]{Colgate1969}
{Colgate} S.~A.,  {McKee} C.,  1969, \apj, 157, 623

\bibitem[\protect\citeauthoryear{{de O{\~n}a-Wilhelmi}, {Rudak}, {Barrio},
  {Contreras}, {Gallant}, {Hadasch}, {Hassan}, {Lopez}, {Mazin}, {Mirabal},
  {Pedaletti}, {Renaud}, {de los Reyes}, {Torres} \& {CTA Consortium}}{{de
  O{\~n}a-Wilhelmi} et~al.}{2013}]{de-Ona-Wilhelmi:2013}
{de O{\~n}a-Wilhelmi} E.,  {Rudak} B.,  {Barrio} J.~A.,  {Contreras} J.~L.,
  {Gallant} Y.,  {Hadasch} D.,  {Hassan} T.,  {Lopez} M.,  {Mazin} D.,
  {Mirabal} N.,  {Pedaletti} G.,  {Renaud} M.,  {de los Reyes} R.,  {Torres}
  D.~F.,    {CTA Consortium} 2013, Astroparticle Physics, 43, 287

\bibitem[\protect\citeauthoryear{{Del Zanna}, {Volpi}, {Amato} \&
  {Bucciantini}}{{Del Zanna} et~al.}{2006}]{Del-Zanna:2006}
{Del Zanna} L.,  {Volpi} D.,  {Amato} E.,    {Bucciantini} N.,  2006, \aap,
  453, 621

\bibitem[\protect\citeauthoryear{{Faucher-Gigu{\`e}re} \&
  {Kaspi}}{{Faucher-Gigu{\`e}re} \& {Kaspi}}{2006}]{Faucher2006}
{Faucher-Gigu{\`e}re} C.-A.,  {Kaspi} V.~M.,  2006, \apj, 643, 332

\bibitem[\protect\citeauthoryear{{Gaensler} \& {Slane}}{{Gaensler} \&
  {Slane}}{2006}]{Gaensler2006}
{Gaensler} B.~M.,  {Slane} P.~O.,  2006, \araa, 44, 17

\bibitem[\protect\citeauthoryear{{Gelfand}, {Slane} \& {Zhang}}{{Gelfand}
  et~al.}{2009}]{Gelfand2009}
{Gelfand} J.~D.,  {Slane} P.~O.,    {Zhang} W.,  2009, \apj, 703, 2051

\bibitem[\protect\citeauthoryear{{H.~E.~S.~S.~Collaboration}, {Abdalla},
  {Abramowski}, {Aharonian}, {Benkhali}, {Ang{\"u}ner}, {Arakawa}, {Arrieta},
  {Aubert}, {Backes} \& et al.}{{H.~E.~S.~S.~Collaboration}
  et~al.}{2018}]{HESS-GPS-2018}
{H.~E.~S.~S.~Collaboration} {Abdalla} H.,  {Abramowski} A.,  {Aharonian} F.,
  {Benkhali} F.~A.,  {Ang{\"u}ner} E.~O.,  {Arakawa} M.,  {Arrieta} M.,
  {Aubert} P.,  {Backes} M.,    et al. 2018, \aap, 612, A1

\bibitem[\protect\citeauthoryear{{Jun}}{{Jun}}{1998}]{Jun1998}
{Jun} B.-I.,  1998, \apj, 499, 282

\bibitem[\protect\citeauthoryear{{Klepser}, {Carrigan}, {de O{\~n}a Wilhelmi},
  {Deil}, {F{\"o}rster}, {Marandon}, {Mayer}, {Stycz}, {Valerius} \&
  {H.~E.~S.~S. Collaboration}}{{Klepser} et~al.}{2013}]{Klepser:2013}
{Klepser} S.,  {Carrigan} S.,  {de O{\~n}a Wilhelmi} E.,  {Deil} C.,
  {F{\"o}rster} A.,  {Marandon} V.,  {Mayer} M.,  {Stycz} K.,  {Valerius} K.,
   {H.~E.~S.~S. Collaboration} 2013, in International Cosmic Ray Conference
  Vol.~33 of International Cosmic Ray Conference, {A Population of
  Teraelectronvolt Pulsar Wind Nebulae in the H.E.S.S. Galactic Plane Survey}.
p.~971

\bibitem[\protect\citeauthoryear{{Kolb}, {Blondin}, {Slane} \& {Temim}}{{Kolb}
  et~al.}{2017}]{Kolb:2017}
{Kolb} C.,  {Blondin} J.,  {Slane} P.,    {Temim} T.,  2017, \apj, 844, 1

\bibitem[\protect\citeauthoryear{{Komissarov}}{{Komissarov}}{2004}]{Komissarov:2004}
{Komissarov} S.~S.,  2004, \mnras, 350, 427

\bibitem[\protect\citeauthoryear{{Mart{\'{\i}}n}, {Torres} \&
  {Pedaletti}}{{Mart{\'{\i}}n} et~al.}{2016}]{Martin2016}
{Mart{\'{\i}}n} J.,  {Torres} D.~F.,    {Pedaletti} G.,  2016, \mnras, 459,
  3868

\bibitem[\protect\citeauthoryear{{Mart{\'{\i}}n}, {Torres} \&
  {Rea}}{{Mart{\'{\i}}n} et~al.}{2012}]{Martin2012}
{Mart{\'{\i}}n} J.,  {Torres} D.~F.,    {Rea} N.,  2012, \mnras, 427, 415

\bibitem[\protect\citeauthoryear{{Olmi} \& {Bucciantini}}{{Olmi} \&
  {Bucciantini}}{2019a}]{Olmi:2019}
{Olmi} B.,  {Bucciantini} N.,  2019a, \mnras, 484, 5755

\bibitem[\protect\citeauthoryear{{Olmi} \& {Bucciantini}}{{Olmi} \&
  {Bucciantini}}{2019b}]{Olmi:2019a}
{Olmi} B.,  {Bucciantini} N.,  2019b, \mnras, 488, 5690

\bibitem[\protect\citeauthoryear{{Olmi}, {Del Zanna}, {Amato}, {Bucciantini} \&
  {Mignone}}{{Olmi} et~al.}{2016}]{Olmi:2016}
{Olmi} B.,  {Del Zanna} L.,  {Amato} E.,  {Bucciantini} N.,    {Mignone} A.,
  2016, Journal of Plasma Physics, 82, 635820601

\bibitem[\protect\citeauthoryear{{Pacini} \& {Salvati}}{{Pacini} \&
  {Salvati}}{1973}]{Pacini:1973}
{Pacini} F.,  {Salvati} M.,  1973, \apj, 186, 249

\bibitem[\protect\citeauthoryear{{Porth}, {Komissarov} \& {Keppens}}{{Porth}
  et~al.}{2014}]{Porth:2014}
{Porth} O.,  {Komissarov} S.~S.,    {Keppens} R.,  2014, \mnras, 438, 278

\bibitem[\protect\citeauthoryear{{Potter}, {Staveley-Smith}, {Reville}, {Ng},
  {Bicknell}, {Sutherland} \& {Wagner}}{{Potter} et~al.}{2014}]{Potter2014}
{Potter} T.~M.,  {Staveley-Smith} L.,  {Reville} B.,  {Ng} C.~Y.,  {Bicknell}
  G.~V.,  {Sutherland} R.~S.,    {Wagner} A.~Y.,  2014, \apj, 794, 174

\bibitem[\protect\citeauthoryear{{Reynolds} \& {Chevalier}}{{Reynolds} \&
  {Chevalier}}{1984a}]{Reynolds_Chevalier84a}
{Reynolds} S.~P.,  {Chevalier} R.~A.,  1984a, \apj, 278, 630

\bibitem[\protect\citeauthoryear{{Reynolds} \& {Chevalier}}{{Reynolds} \&
  {Chevalier}}{1984b}]{Reynolds1984}
{Reynolds} S.~P.,  {Chevalier} R.~A.,  1984b, \apj, 278, 630

\bibitem[\protect\citeauthoryear{{Sedov}}{{Sedov}}{1959}]{Sedov1959}
{Sedov} L.~I.,  1959, {Similarity and Dimensional Methods in Mechanics}

\bibitem[\protect\citeauthoryear{{Tanaka} \& {Takahara}}{{Tanaka} \&
  {Takahara}}{2010}]{Tanaka2010}
{Tanaka} S.~J.,  {Takahara} F.,  2010, \apj, 715, 1248

\bibitem[\protect\citeauthoryear{{Temim}, {Slane}, {Kolb}, {Blondin}, {Hughes}
  \& {Bucciantini}}{{Temim} et~al.}{2015}]{Temim2015}
{Temim} T.,  {Slane} P.,  {Kolb} C.,  {Blondin} J.,  {Hughes} J.~P.,
  {Bucciantini} N.,  2015, \apj, 808, 100

\bibitem[\protect\citeauthoryear{{Torres}}{{Torres}}{2017a}]{Torres2017b}
{Torres} D.~F.,  2017a, {Modelling Pulsar Wind Nebulae}.
Vol.~446

\bibitem[\protect\citeauthoryear{{Torres}}{{Torres}}{2017b}]{Torres2017}
{Torres} D.~F.,  2017b, \apj, 835, 54

\bibitem[\protect\citeauthoryear{{Torres}}{{Torres}}{2018}]{Torres2018}
{Torres} D.~F.,  2018, Nature Astronomy, 2, 247

\bibitem[\protect\citeauthoryear{{Torres}, {Cillis}, {Mart{\'{\i}}n} \& {de
  O{\~n}a Wilhelmi}}{{Torres} et~al.}{2014}]{Torres2014}
{Torres} D.~F.,  {Cillis} A.,  {Mart{\'{\i}}n} J.,    {de O{\~n}a Wilhelmi} E.,
   2014, Journal of High Energy Astrophysics, 1, 31

\bibitem[\protect\citeauthoryear{{Torres} \& {Lin}}{{Torres} \&
  {Lin}}{2018}]{Torres2018b}
{Torres} D.~F.,  {Lin} T.,  2018, \apjl, 864, L2

\bibitem[\protect\citeauthoryear{{Torres}, {Lin} \& {Coti Zelati}}{{Torres}
  et~al.}{2019}]{Torres2019}
{Torres} D.~F.,  {Lin} T.,    {Coti Zelati} F.,  2019, \mnras, 486, 1019

\bibitem[\protect\citeauthoryear{{Truelove} \& {McKee}}{{Truelove} \&
  {McKee}}{1999}]{Truelove1999}
{Truelove} J.~K.,  {McKee} C.~F.,  1999, \apjs, 120, 299

\bibitem[\protect\citeauthoryear{{van der Swaluw}, {Achterberg}, {Gallant},
  {Downes} \& {Keppens}}{{van der Swaluw} et~al.}{2003}]{van-der-Swaluw:2003}
{van der Swaluw} E.,  {Achterberg} A.,  {Gallant} Y.~A.,  {Downes} T.~P.,
  {Keppens} R.,  2003, \aap, 397, 913

\bibitem[\protect\citeauthoryear{{van der Swaluw}, {Achterberg}, {Gallant} \&
  {T{\'o}th}}{{van der Swaluw} et~al.}{2001}]{van-der-Swaluw:2001}
{van der Swaluw} E.,  {Achterberg} A.,  {Gallant} Y.~A.,    {T{\'o}th} G.,
  2001, \aap, 380, 309

\bibitem[\protect\citeauthoryear{{Vorster}, {Tibolla}, {Ferreira} \&
  {Kaufmann}}{{Vorster} et~al.}{2013}]{Vorster2013}
{Vorster} M.~J.,  {Tibolla} O.,  {Ferreira} S.~E.~S.,    {Kaufmann} S.,  2013,
  \apj, 773, 139

\end{thebibliography}


\section*{Data Availability}

The data for the models underlying this article will be shared on reasonable request to the corresponding author.

\appendix

\section{Profiles for the shocked ejecta}\label{sec:a1}

Here we will recall the \citet{Sedov1959} results, used here in the \citet{Bandiera1984} formulation to describe the shocked ejecta profiles.
We have considered only the case of a constant ambient medium, namely with $\rho=\rhoism$.
The boundary conditions behind the forward shock are
\begin{eqnarray}
\rho_2 & = & \frac{\gamma+1}{\gamma-1} \rhoism \\
v_2 & = & \frac{2}{\gamma+1} \dotRforw \\
P_2 & = & \frac{2}{\gamma+1} \rhoism \dotRforw^2.
\end{eqnarray}
being $\Rforw$ the SNR FS radius and $\dotRforw$ its expansion velocity. In our approach, we take these two latter values from \citet{Truelove1999} shock trajectories.
We factorize the values of the profiles to the boundary conditions such that
\begin{eqnarray}
\rhoej(\xi) & = & \bar{\rho}(\xi)\, \rho_2(t) \\
\vej(\xi) & = & \bar{v}(\xi)\, v_2(t) \\
\Pej(\xi) & = & \bar{P}(\xi)\, P_2(t)
\end{eqnarray}
where $\xi=\Rpwn/\Rforw$. The $\bar{\rho}$, $\bar{v}$ and $\bar{P}$ coefficients are given by
\begin{eqnarray}
\bar{\rho}(\xi) & = & B^{1-2(\gamma-1)a} C^{3c-1} D^{3b-1} \\
\bar{v}(\xi) & = & \tilde{v}\,\xi \\
\bar{P}(\xi) & = & A^{\frac{6}{5}} C^{3c} D^{1+b}
\end{eqnarray}
with
\begin{eqnarray}
A & = & \frac{\tilde{v}}{D} \\
B & = & \frac{2\gamma}{\gamma-1}\left(\tilde{v}-\frac{\gamma+1}{2\gamma} \right)/D \\
C & = & \frac{2}{\gamma-1}\left(\frac{\gamma+1}{2}-\tilde{v} \right)/D \\
D & = & \frac{2(3\gamma-1)}{7-\gamma} \left(\frac{\gamma+1}{2}\frac{5}{3\gamma-1} -\tilde{v}\right) \\
a & = & \frac{1}{2\gamma+1} \\
b & = & \frac{\gamma+1}{3\gamma-1} \\
c & = & -\frac{\gamma}{3(2-\gamma)}.
\end{eqnarray}
Additionally, we have the following expression for $\xi$
\begin{equation}
\xi=A^{-\frac{2}{5}} B^{(\gamma-1)a} D^{-b}
\end{equation}
which is needed to find the value for $\tilde{v}$ using an iterative method. 
Once we get the value for $\tilde{v}$, we calculate all the coefficients and the values for $\rhoej$, $\vej$ and $\Pej$ at the position of the PWN radius.

\label{lastpage}
\end{document}